\documentclass[%
reprint,
groupedaddress,
 amsmath,amssymb,
 aps,
pra,
]{revtex4-1}

\usepackage{graphicx}
\usepackage{dcolumn}
\usepackage{bm}

\usepackage{braket}

\usepackage{xcolor}

\begin{document}


\title{Entropic Quantum Machine}

\author{Ryoko Hatakeyama}
\author{Akira Shimizu}%
 \email{Corresponding author: shmz@as.c.u-tokyo.ac.jp}
\affiliation{Komaba Institute for Science, The University of Tokyo, 3-8-1 Komaba, Meguro, Tokyo 153-8902, Japan}
\affiliation{Department of Basic Science, The University of Tokyo, 3-8-1 Komaba, Meguro, Tokyo 153-8902, Japan}




\date{\today}

\begin{abstract}
We study nanomachines whose relevant (effective) degrees of freedom $f \gg 1$
but smaller than $f$ of proteins.
In these machines, 
both the entropic and the quantum effects 
over the whole system play the essential roles
in producing nontrivial functions.
We therefore call them entropic quantum machines (EQMs).
We propose a systematic protocol for designing the EQMs,
which enables 
a rough sketch, accurate design of equilibrium states, 
and accurate estimate of response time.
As an illustration, we design a novel EQM, which
shows two characteristic shapes.
One can switch from one shape to the other
by changing temperature or by applying a pulsed external field.
We discuss two potential applications of this example of an EQM.
\end{abstract}

\maketitle


\section{introduction}
Along with the rapid development of nanomachines \cite{?ki1976, Iwamura1988, Rebek1979, Shinkai1980, Anelli1991, Bissell1994, Kelly1999, Koumura1999, Klok2008, Eelkema2006, Wang2011, Hernandez2004, Leigh2003, VonDelius2010, Barrell2011, Shirai2005, Collier1999, Green2007, Bottari2003, Perez2004, DeBo2018, Hanggi2009, Kupchenko2011, Maksymovych2005, Yamaki2007a, Vacek2007, Baudry2006, Arkhipov2006, Fendrich2006, Strambi2010, Horodecki2013, Skrzypczyk2014, Cai2010, Malabarba2015, Hofer2017, Hummer2001, Hospital2015},
methods of designing them 
have been attracting much attention \cite{Hanggi2009, Kupchenko2011, Maksymovych2005, Yamaki2007a, Vacek2007, Baudry2006, Arkhipov2006, Fendrich2006, Strambi2010, Horodecki2013, Skrzypczyk2014, Cai2010, Malabarba2015, Hofer2017}.
Although both the quantum and the entropic effects should be taken into account, in general, 
to design nanomachines, 
either one may be neglected or simplified in some cases.
For example, 
when the relevant (i.e., effective) degrees of freedom $f$ of a nanomachine is small, 
i.e. $f \sim 1$ as in Refs.~\cite{Yamaki2007a, Hofer2017}, 
one can neglect the entropic effect
(apart from that of reservoirs).
In this case, however, only a simple function is expected
because only the quantum effect with small degrees of freedom is 
available.
By contrast, 
when $f$ 
is large, as in proteins, 
more complex functions are expected
that utilize the entropic effect as well.
To design the machine in this case, it is customary to calculate the free energy 
of a {\em classical} model 
\cite{Vacek2007, Baudry2006, Arkhipov2006, Hummer2001, Hospital2015}, 
where the quantum effect is considered 
only {\em locally} to determine the model parameters (such as the spring constant)
in the classical model.
This type of approach was taken also for analyzing DNA \cite{Kittel1969, Holubec2012}.

Then, 
let us consider nanomachines whose $f$ takes an intermediate value, 
$f \gg 1$ but smaller than $f$ of proteins.
Since $f \gg 1$, 
they can have more complicated functions 
than the machines with $f \sim 1$.
In particular, they can utilize the 
{\em entropic effect}
to realize functions.
At the same time,
since $f$ is smaller than that of proteins,
the nanomachines can utilize the 
{\em quantum effect over the whole machine}.
This suggests that 
the machines can be smaller than a protein
that has the same function.
For these reasons, 
such nanomachines seem very interesting.
We call them `entropic quantum machines' (EQMs).

However, 
none of the previous methods that are mentioned above 
are applicable to quantitative design of EQMs
because 
both the entropic effect 
and 
the quantum effect over the whole machine
should be taken into account.
A possible approximate method is the 
density functional method \cite{Hohenberg1964, Burke2012}.
However,
for nontrivial quantum systems 
such as the frustrated many-body systems,
its accuracy is generally insufficient,
and other elaborate methods \cite{Jeckelmann2002, Hallberg1995, Weichselbaum2009, Schollwock2011, Shibata2003, Foulkes2001, Gohlke2017, Sugiura2012, Sugiura2013, Hyuga2014} are usually employed.
Since nontrivial quantum systems will be appropriate for EQMs, 
it seems better to adopt 
such elaborate methods.
However, these methods
focused mainly on the analyses of properties of {\em given} systems.
In order to design a new nanomachine,
one should also be able to {\em sketch 
the system itself}
before analyzing its properties in detail.

In this paper, we propose 
a systematic protocol for designing the EQMs.
It consists of three steps,
1: sketch the system itself,
2: 
optimize the values of the parameters,
and 
3: obtain the response time of the EQM.
As an illustration, we design a novel EQM, which
shows two characteristic shapes (particle distributions).
One can switch from one shape to the other
by changing temperature or by applying a pulsed external field.
We discuss two potential applications of this EQM.
One is to control reaction between a receptor 
and an agonist.
The other is to work as a nanozyme, using which 
one can choose between two different reactions to catalyze.

\section{Protocol for designing EQMs}\label{sec:recipe}

We focus on EQMs that operate not 
by chemical reactions 
but by physical stimuli such as an external field and temperature
change.

To design such EQMs, 
we make a full use of the thermal pure quantum (TPQ) formulation \cite{Sugiura2012, Sugiura2013, Hyuga2014, sugiura2017formulation}.
The TPQ formulation is a full reformulation, based on the pure state statistical mechanics, of quantum statistical mechanics.  
It represents any equilibrium state by a single state vector, 
called a TPQ state, 
{\em without} introducing any ancilla systems (such as a reservoir).
It was proved rigorously that 
one can obtain all statistical-mechanical quantities
from a single TPQ state.
Both the entropic and the quantum effects 
can be accurately calculated,
with exponentially small errors.

Many useful formulas were developed, including the one by which 
the thermodynamic functions are obtained accurately from the norm of a single TPQ state. 
The TPQ formulation is not only interesting from a fundamental viewpoint
but also useful for practical calculations because 
it gives accurate results for any quantum many-body systems whose size 
is too large for numerical diagonalization.
Noting this advantage, 
we propose the following procedure for designing EQMs.

{\em Step 1.}
By qualitative and semi-quantitative considerations, 
sketch an EQM according to the purpose.
Utilize, for example, 
competing terms in the Hamiltonian
and a large degree of degeneracy,  
as we will illustrate in the next section.

{\em Step 2.}
Calculate equilibrium properties of the EQM at various temperatures using the 
TPQ formulation.
This enables one 
to confirm the expected properties
and 
to optimize 
the values of the parameters in the Hamiltonian.

{\em Step 3.}
Calculate the time evolution of an initial TPQ (equilibrium) state after the 
quench, i.e., after the application of an external 
field. 
Confirm that the final stationary state agrees with the equilibrium state
of the same energy, which is obtained in Step 2. 
This enables one 
to obtain the response time of the EQM
and
to determine appropriate values of the height and width of 
the pulse of the external field.

In the following sections, 
we explain the above protocol in detail by showing an example.

\section{Step 1: Sketch of an EQM}

\subsection{Proposed system}
\begin{figure*}
    \centering
    \includegraphics[width=\textwidth]{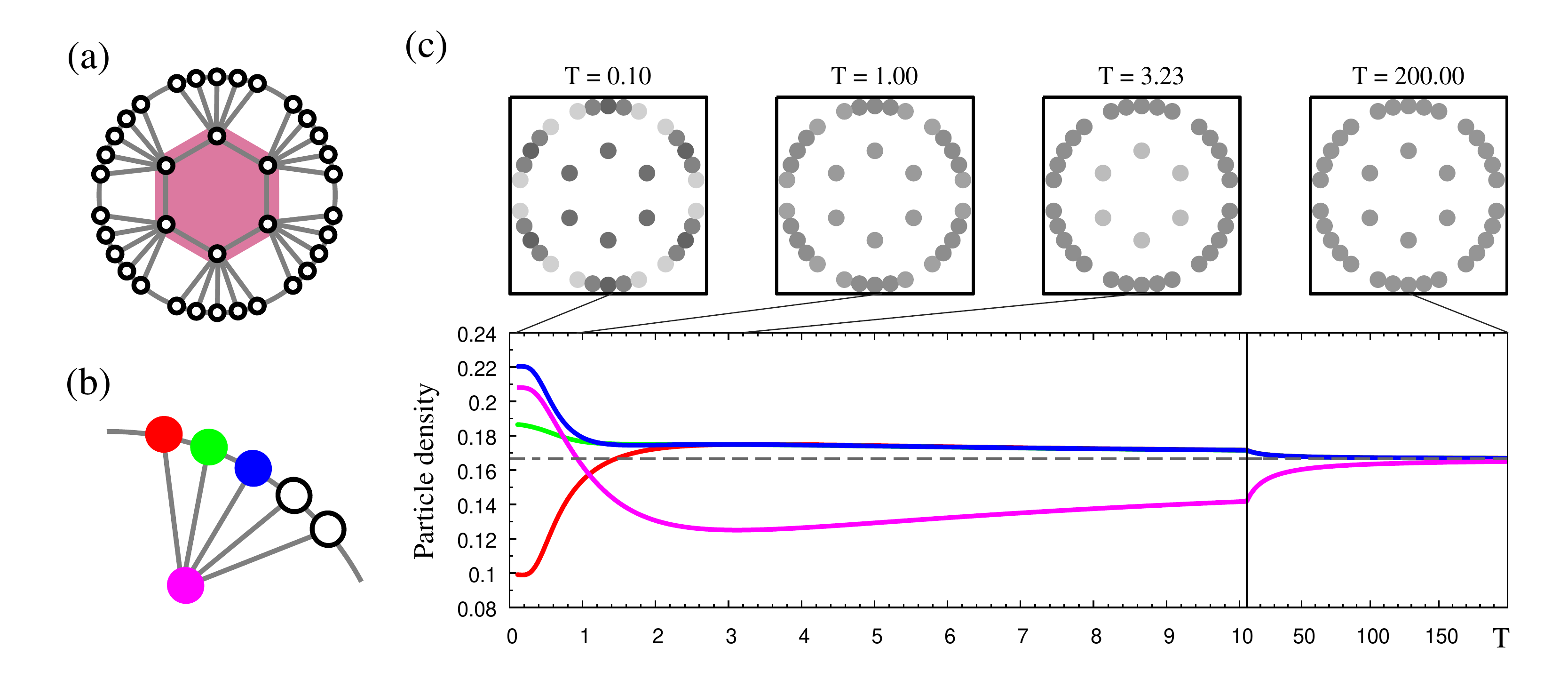}
\caption{
(a) Proposed EQM. 
The dots and the line segments represent the sites and the bonds of the lattice respectively.
When an external field is applied, it is applied on the purple area.
(b) A part of the EQM.
(c) Particle distribution as a function of temperature, $T$.
The dashed line represents $n_{\rm av}$.
The scale of $T$ is in units of $J$.
}
    \label{fig:equilibrium}
\end{figure*}

From the sketchy considerations which will be described shortly, 
we find that the following system exhibits two characteristic shapes.
It is a system of particles (spinless fermions or hardcore bosons)
on the double-circle lattice shown in Fig.~\ref{fig:equilibrium} (a).
Such a system seems implementable experimentally in various systems, 
such as quantum-dot arrays, optical lattices and large molecules.
As its natural Hamiltonian we assume 
\begin{equation}
H=H_{\mathrm{hop}} + H_{\mathrm{rep}}, 
\end{equation}
where 
\begin{align}
H_{\mathrm{hop}} &= - J {\sum}' c_i^\dagger c_j + \mathrm{h. c.},
\\
H_{\mathrm{rep}} &= V {\sum}' n_i n_j.
\label{eq:hamiltonian}
\end{align}
Here,  
$c_i$ annihilates a particle on site $i$, 
$n_i = c_i^\dagger c_i$, 
$J$ ($\geq 0$)  is the hopping energy,  
$V$ ($\geq 0$)  is the repulsion between two adjacent sites,
and ${\sum}'$ denotes the sum over pairs of sites connected by a bond.
We here take $J$ and $V$ common to all bonds.
Hence, $H$ essentially has {\em only a single parameter $V/J$}.
Multiplying $J$ and $V$ simultaneously by the same factor 
results only in change of the scales of temperature and time by that factor.
By contrast, 
a machine that has many microscopic parameters is
neither feasible nor uninteresting 
because nontrivial behaviors are obviously expected by fine tuning of 
such many parameters.

We hereafter use $J$ and $\hbar/J$ as the units of temperature and time, 
respectively. 


In this system, 
$H_{\mathrm{hop}}$ and $H_{\mathrm{rep}}$
compete with each other.
They induce the wave and particle natures, respectively,
as will be discussed in Secs.~\ref{sec:Nontrivial_switching}
and \ref{sec:qande}.
Such competition between the complementary natures is typical of 
quantum systems.
With increasing temperature $T$, 
energy and entropy 
also compete with each other to minimize the free energy.
By utilizing these competitions, we realize a nontrivial switching of
particle distribution with increasing $T$.

Let $\braket{n_i}_T$ be the particle density at site $i$, 
where $\braket{ \bullet }_T$ denotes the expectation value at temperature $T$.
It is obvious that, 
at $T = \infty$,
the distribution becomes uniform to maximize entropy: 
\begin{equation}
\braket{n_i}_\infty = n_{\rm av} := N/L \ \mbox{for all $i$}.
\end{equation}
Here, $L = 36$ is the number of sites on the lattice and $N$ is the total number of particles, which is assumed to be fixed,
independently of $T$. 
At finite $T$, 
$\braket{n_i}_T$ can differ from site to site, 
but 
it takes the same value, denoted by $\braket{n_{\rm in}}_T$,  
in the inner circle by symmetry.
Starting from low $T$, 
we shall realize a nontrivial switching with increasing $T$,
from $\braket{n_{\rm in}}_T > n_{\rm av}$
to $\braket{n_{\rm in}}_T < n_{\rm av}$
(and finally to $\braket{n_{\rm in}}_T = n_{\rm av}$).
By contrast, in most other systems with a single parameter $V/J$
such a switching is impossible  
because the density distribution at low $T$ just 
approaches monotonically the uniform one with increasing $T$.

\subsection{Nontrivial switching}\label{sec:Nontrivial_switching}
To sketch 
the system for the nontrivial switching, 
we investigate low-temperature states,
assuming $n_{\rm av} \ll 1$ 
so that the particles can hop easily.

When $H=H_{\mathrm{hop}}$ (i.e., $V=0$), 
we expect 
$\braket{n_{\rm in}}_T > n_{\rm av}$ for $T \ll J$ 
because particles in the inner circle can hop to more sites
and gain an energetic benefit.
[This expectation is confirmed in Sec~\ref{sec:qande}.]
This is a result of the wave nature 
and an energetic effect.

When $H=H_{\mathrm{rep}}$ (i.e., $J=0$), on the other hand, 
a ground state is a state such that no particles are adjacent to each other
to reduce the repulsive interaction. 
Since there are many such configurations, 
the ground states are degenerate with a high degree, 
as shown in Table \ref{tbl:Vonly}.
At low temperature such that $T \ll V$,
particles occupy these states with equal weights
to maximize entropy.
The ratio of $\braket{n_{\rm in}}_T$ to $n_{\rm av}$ 
in this case is also shown in the table.
We observe that $\braket{n_{\rm in}}_T < n_{\rm av}$, 
which happens because the outer circle has 
a greater number of possible configurations
than the inner circle.
This is a result of the particle nature and an entropic effect.

\begin{table}
\caption{\label{tbl:Vonly}
Degree of degeneracy of the ground states, 
and the ratio
$\braket{n_{\rm in}}_T/ n_{\rm av}$ for $T \ll V$,
when $H=H_{\mathrm{rep}}$.
}
\begin{ruledtabular}
\begin{tabular}{c|cccccc}
N       & 3     & 4     & 5         & 6         & 7         & 8  
\\
\hline
degeneracy  & 5088  & 29454 & 115320    & 313329    & 596202    & 791664
\\
$\braket{n_{\rm in}}_T/n_{\rm av}$
            & 0.795 & 0.703 & 0.618     & 0.538     & 0.462     & 0.389
\end{tabular}
\end{ruledtabular}
\end{table}

It is seen from these observations that 
the switching 
from $\braket{n_{\rm in}}_T > n_{\rm av}$
to $\braket{n_{\rm in}}_T < n_{\rm av}$ should be possible if 
$H_{\mathrm{hop}}$ (wave nature and energetic effect)
and 
$H_{\mathrm{rep}}$ (particle nature and entropic effect)
play dominant roles at lower and higher $T$, respectively.
This idea is realized as follows.

We utilize the high degeneracy of the ground states of $H_{\mathrm{rep}}$.
For this purpose, we limit ourselves to the manifold of these ground states by taking
\begin{align}
J &\ll V,
\label{eq:J<<V}
\\
T &< V. 
\label{eq:T<V}
\end{align}
Under these conditions, 
we expect the following switching behavior,
as $T$ is increased from $T<J$ to $T>J$.\\
(i) At low temperature $T < J$,
$H_{\mathrm{hop}}$ is significant because it lifts the degeneracy
(whereas $H_{\mathrm{rep}}$ just determines the manifold of the relevant states).
It lowers the energies of states with larger $\braket{n_{\rm in}}_T$
because particles in the inner circle can hop to more sites
and gain an energetic benefit.
Since particles occupy such lower-energy states for $T < J$, 
we expect $\braket{n_{\rm in}}_T > n_{\rm av}$.
Note that this will be more effective for smaller $N$ and $V/J$
because hopping is suppressed for larger $N$ and $V/J$.\\
(ii) At higher temperature $J < T$ ($\ll V$), 
$H_{\mathrm{hop}}$ becomes irrelevant, and 
particles occupy all states of the manifold with almost equal weights.
Consequently, $\braket{n_{\rm in}}_T < n_{\rm av}$ should be realized.
This is more effective for {\em larger} $N$ (as long as $n_{\rm av} \ll 1$), 
as seen from 
Table \ref{tbl:Vonly}, and for larger $V/J$.

We can estimate appropriate values of $N$ and $V/J$ 
from 
the above arguments, as follows.
We have assumed $n_{\rm av} \ll 1$, and took $V/J \gg 1$.
Smaller $N$ and $V/J$ are better for (i), 
whereas larger $N$ and $V/J$ are better for (ii).
Considering this trade-off between (i) and (ii),  
we here take $N=6$ (so that $n_{\rm av}=1/6$) 
and $V/J \simeq 5$.
We will show that the nontrivial switching is indeed realized for this 
choice of $N$ and $V/J$.

\section{Step 2: Quantitative analysis of equilibrium states}
We analyze equilibrium states of the above system
at various temperatures.
For this purpose, we employ the TPQ formulation 
\cite{Sugiura2012, Sugiura2013, Hyuga2014, sugiura2017formulation},
as explained in Sec.~\ref{sec:recipe}.
Using this formulation, 
we can calculate 
both the entropic and the quantum effects 
accurately, with exponentially small errors at any non-vanishing temperature, 
with far less computer resources than the exact diagonalization method.

To be concrete, we assume spinless fermions,
which may be realized, e.g., as spin-polarized electrons.
[We can obtain similar results for hardcore bosons \cite{MT}, which may be realized, e.g.,
in optical lattices.]
To find the optimal value of $V/J$,
we introduce the figure of merit defined as
the smaller one between 
the highest excess density 
$\max_T [\braket{n_{\rm in}}_T - n_{\rm av} ]$
and
the deepest deficient density 
$\max_T [n_{\rm av} - \braket{n_{\rm in}}_T]$
in the inner circle,
and find that 
$V/J = 4.7$ is optimal.
We thus show the results at $V/J = 4.7$ 
in the following.

The calculated distribution of particles is shown in 
Fig.~\ref{fig:equilibrium} (c) as a function of $T$.
Here and after, we present the distribution of only four 
sites that are colored in Fig.~\ref{fig:equilibrium} (b) because the other sites are identical to either of these sites by symmetry. 
At low temperature $T = T_0 := 0.1$, 
we observe that the inner circle has higher density 
$\braket{n_{\rm in}}_T = 0.208 > n_{\rm av} = 0.166 \cdots$, 
as expected from (i) above.
We also find that the particle distribution
shows a characteristic pattern in the outer circle
(whereas the distribution is always  uniform in the inner circle by symmetry).
This pattern is formed principally by $H_{\mathrm{hop}}$ 
and will be useful for certain applications \cite{MT}.

At higher temperature 
$T= T_* :=3.23$, for which $J<T<V$,
we find that $\braket{n_{\rm in}}_T$ takes the minimum value 
$\braket{n_{\rm in}}_T = 0.125 < n_{\rm av}$,
as expected from (ii) above.
In this case, the particle distribution in the outer circle is almost uniform,
as in the inner circle,
because $H_{\mathrm{hop}}$ is irrelevant at this temperature.
We have also found that the entropy
increases quickly with increasing $T$, 
and, at $T=T_*$, 
it grows to almost 88\% 
of the total value 
$S_{\rm total} = \ln {36 \choose 6}$ \cite{MT}.
This confirms (ii), i.e.,
$\braket{n_{\rm in}}_T < n_{\rm av}$ is realized by the entropic effect.

At an intermediate temperature $T=T_{\rm m} :=1.00$,
we observe that the particle distribution is almost uniform all over the system, 
$\braket{n_i}_T \simeq n_{\rm av}$ for all $i$,
because of an interplay of the two effects discussed in (i) and (ii).
At very high temperature $T \gg V$, 
it is obvious that $\braket{n_i}_T \simeq \braket{n_i}_\infty = n_{\rm av}$ for all $i$,
as shown in Fig.~\ref{fig:equilibrium}(c) for $T=200$.

We have thus realized the switching 
from $\braket{n_{\rm in}}_T > n_{\rm av}$, 
to the nearly uniform distribution, 
to $\braket{n_{\rm in}}_T < n_{\rm av}$,
and finally to the uniform distribution.


\section{Step 3: Response to external field}

\subsection{Thermalization}
The above switching is realized by increasing $T$.
There are various methods for increasing $T$, 
such as the heat contact with a hot reservoir.
It has been recently clarified that,
when an external field is applied to a quantum many-body system, 
it approaches a new equilibrium state, i.e. `thermalizes,'
even if the system is completely isolated from other systems,
provided that the Hamiltonian is natural enough \cite{Neumann1929, Berry1977, Trotzky2012, Deutsch1991, Srednicki1994, Tasaki1998, Rigol2008, DAlessio2016}. 
Since heat exchange with a reservoir is unnecessary, 
we expect that the thermalization 
enables the system to reach a hot equilibrium state 
faster than the heat contact.
We therefore study whether our EQM thermalizes,
how quick it is, 
and what profile (spatial and temporary) of the external field is appropriate.
This provides us with a way of fast switching  of the EQM. 




\subsection{Method of calculating dynamical properties}
For this purpose, 
we employ the recent method \cite{Endo2018} for 
analyzing dynamical properties of the TPQ states.


The initial state is taken as the equilibrium state 
of temperature $T_0=0.1$,
for which $\braket{n_{\rm in}}_{T_0} > n_{\rm av}$ as shown above.
Suppose that a pulsed external field $h$, 
shown in Fig.~\ref{fig:evolution} (a),
is applied to the system,
in order to feed energy.
We consider the case where 
$h$ is applied on the purple area of Fig.~\ref{fig:equilibrium} (a).
We assume that $h$ interacts with the system via
\begin{align}
        H_{\mathrm{ext}} = h {\sum}^{\prime \prime} n_i,
        \label{eq:external field}
\end{align}
where ${\sum}^{\prime \prime}$ denotes the sum over the sites of the inner circle.

The initial equilibrium state is taken as the TPQ state 
$\ket{T_0}$ \cite{Sugiura2013}.
It was shown rigorously that
$\ket{T_0}$ gives 
the same time evolution of statistical-mechanical observables
as the Gibbs state, 
with an exponentially small error \cite{Endo2018}.
Since $\ket{T_0}$ is a pure quantum state, 
its time evolution can be calculated  
with far less computational resources than that of the Gibbs state, which is a mixed quantum state.
We use the Chebyshev-polynomial expansion \cite{Tal-Ezer1984} 
for the time-evolution operator $U(t)$, where $t$ is time.
The end point $t_{\rm end}$ is taken 
sufficiently longer than the relaxation time $\tau$ (see below).
We further calculate the round-trip evolution 
\begin{align}
\ket{\tilde{T_0}} := 
U(-t_{\rm end}) U(t_{\rm end}) \ket{T_0}, 
\end{align}
which should equal $\ket{T_0}$ 
if the time evolution is correctly carried out. 
By taking the Chebyshev polynomials up to the $671$st order at every $\Delta t=10$, 
we obtain the fidelity 
\begin{align}
\frac{|\braket{T_0 |\tilde{T_0}}|^2}
{\braket{T_0 | T_0}\braket{\tilde{T_0}|\tilde{T_0}}}
\end{align}
as high as
$1 \pm 1.0 \times 10^{-8}$. 
This confirms 
the accuracy of our time evolution.

%

\subsection{Results}
Our purpose is to increase $T$ from $T_0$ to $T_*$ 
by applying $h$.
We therefore take 
the magnitude $h$ and pulse width $t_{\rm pulse}$
in such a way that 
the energy increase by $h$ agrees with 
the energy difference 
(which is calculated using the TPQ formulation)
between the equilibrium states at these temperatures. 
We thus take $h=20$ and $t_{\rm pulse} = 0.1188$.

\begin{figure*}
    \centering
    \includegraphics[width=\textwidth]{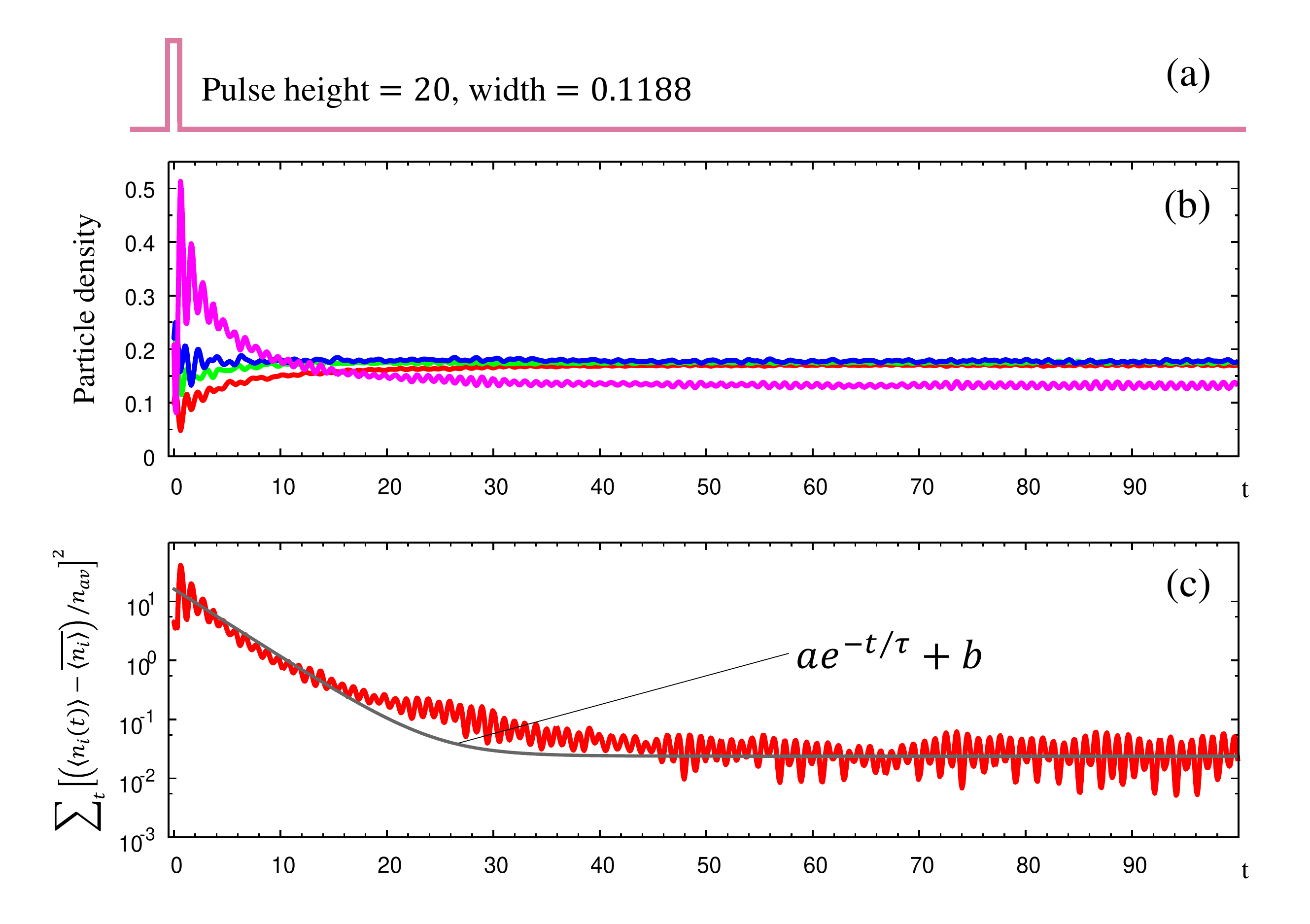}
\caption{
Time evolution of 
(a) external field, 
(b) $\braket{n_i(t)}$ for the 
four sites shown in Fig.~\ref{fig:equilibrium}(b),
and 
(c) $\sum_i [(\braket{n_i(t)}- \overline{\braket{n_i}})/n_{\rm av}]^2$.
The scale of time $t$ is in units of $\hbar/J$.
}
    \label{fig:evolution}
\end{figure*}

Figure \ref{fig:evolution}(b) shows 
the time evolution of $\braket{n_i(t)}$ for the 
four sites shown in Fig.~\ref{fig:equilibrium}(b).
It is seen that all $\braket{n_i(t)}$ approaches a stationary value,
defined by the time average $\overline{\braket{n_i}}$
over the interval $[t_{\rm end} - 20, t_{\rm end}]$,
apart from small fluctuation.
To confirm that $\overline{\braket{n_i}}$ agrees with 
the equilibrium value $\braket{n_i}_{T_*}$,
we calculate 
\begin{align}
\sum_i 
[(\overline{\braket{n_i}}- \braket{n_i}_{T_*})/n_{\rm av}]^2,
\end{align}
and find it as small as $0.0234$.
Since 
such a small deviation is normally 
observed in thermalization of isolated systems of finite size,
we conclude that 
the equilibrium values are well realized.

To estimate the response time,
we also calculate 
\begin{align}
\sum_i 
[(\braket{n_i(t)}- \overline{\braket{n_i}})/n_{\rm av}]^2,
\end{align}
which is plotted in Fig.~\ref{fig:evolution}(c). 
It is well fitted by $a e^{-t/\tau} + b$ (gray solid line)
with $\tau=3.777$, $a=16.27$ and $b=0.02381$. 
Since this $\tau$ is the same order of magnitude as the
characteristic time scale $\hbar/J$ of the system,
the response is fast enough.

%

\section{Quantum and entropic effects}
\label{sec:qande}

We have explained our protocol for designing EQMs by showing an example.
Before presenting its possible applications, 
we discuss interplay of the quantum and entropic effects in this EQM
by studying the two limiting cases where 
(a) only the hopping term exists ($J=1$, $V=0$) and 
(b) only the repulsion term exists ($J=0$, $V=4.7$).

\begin{figure*}
    \centering
\includegraphics[width=\textwidth]{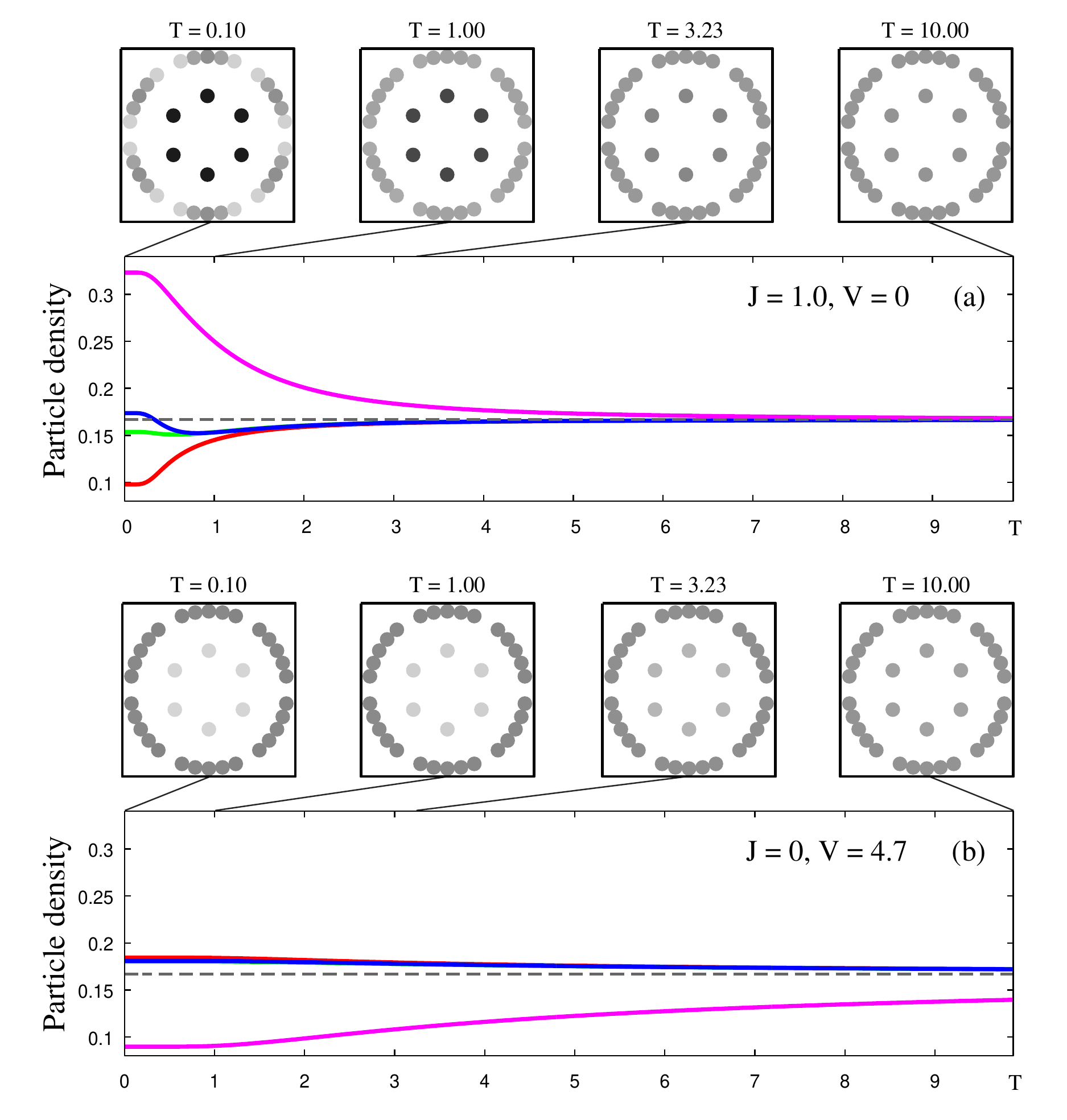}
    \caption{
    Particle distribution as a function of $T$ for (a) $V=0$ and (b) $J=0$.
    The dashed lines represent $n_{\rm av}$.
The scale of $T$ is taken as the same as in Fig.~\ref{fig:equilibrium}.
    }
    \label{fig:J=0orV=0}
\end{figure*}

In Fig.~\ref{fig:J=0orV=0}, 
we show the distribution of particles in the two cases 
as a function of $T$.
The scale of $T$ is taken as the same as in Fig.~\ref{fig:equilibrium}
for the sake of comparison.

In case (a), 
the particles form the Fermi sea at low temperature.
The sea is composed of single-particle wavefunctions which are coherent all over the system. 
A single-particle wavefunction tends to have lower energy when it has 
larger magnitude in the inner circle
because more hoppings are available for the hopping paths that visit the inner circle, which results in constructive interference in the inner circle.
Consequently, the inner circle has higher particle density 
than the outer circle
at low temperature, $T < J =1$.

Note that this is a benefit of the wave nature of quantum mechanics. 
In fact, in order to mimic this density distribution 
using a {\em classical} system, 
one has to increase the number of parameters in such a way that the sites in the inner circle have lower site energies.
By contrast, our quantum system essentially has 
only a single parameter $V/J$, and all sites have the same energy 
(taken $0$).
Nevertheless, the quantum interference effect\footnote{
By the quantum interference we mean the interference between different paths of particles, which is caused by the hopping term in the Hamiltonian.  
}
 yields the 
higher density in the inner circle at low temperature such as $T=0.1$.

In case (b), 
such a wave nature is lost and the particles can be regarded 
as classical particles. 
This particle nature results in {\em lower} density in the inner circle
at low temperature, $T < V = 4.7$,
due to the entropic effect
because the outer circle has a greater number of possible configurations 
than the inner circle.

We have thus confirmed that the quantum and the entropic effects give the opposite 
effects on the particle distribution at low temperature.
Comparing these results with those of Fig.~\ref{fig:equilibrium}, 
we conclude that the nontrivial switching of the particle distribution 
of our EQM is realized as a result of competition between 
the hopping and the repulsion terms in the Hamiltonian, 
i.e., 
between the quantum and the entropic effects.

Note also that the particle distribution 
in the {\em outer} circle is nonuniform 
in Fig.~\ref{fig:J=0orV=0} (a).
This is also a quantum interference effect, i.e., a manifestation of the wave nature.
[In fact, 
the distribution becomes uniform in Fig.~\ref{fig:J=0orV=0} (b),
where particles can be regarded as classical ones.]
This quantum interference effect survives in Fig.~\ref{fig:equilibrium}.

\section{Potential applications of the proposed EQM}

We have demonstrated an example of EQM, which exhibits nontrivial 
changes in the particle distribution.
As discussed in Sec.~\ref{sec:qande}, such a property 
is realized by utilizing both the quantum and entropic effects, 
despite the fact that our EQM has essentially a single 
parameter, $V/J$.
We here discuss its potential applications.

Note that, in these applications, 
we can change the state of the EQM {\em reversibly and repeatably}.

\subsection{Control of agonist}
\begin{figure*}
\centering
\includegraphics[width=\textwidth]{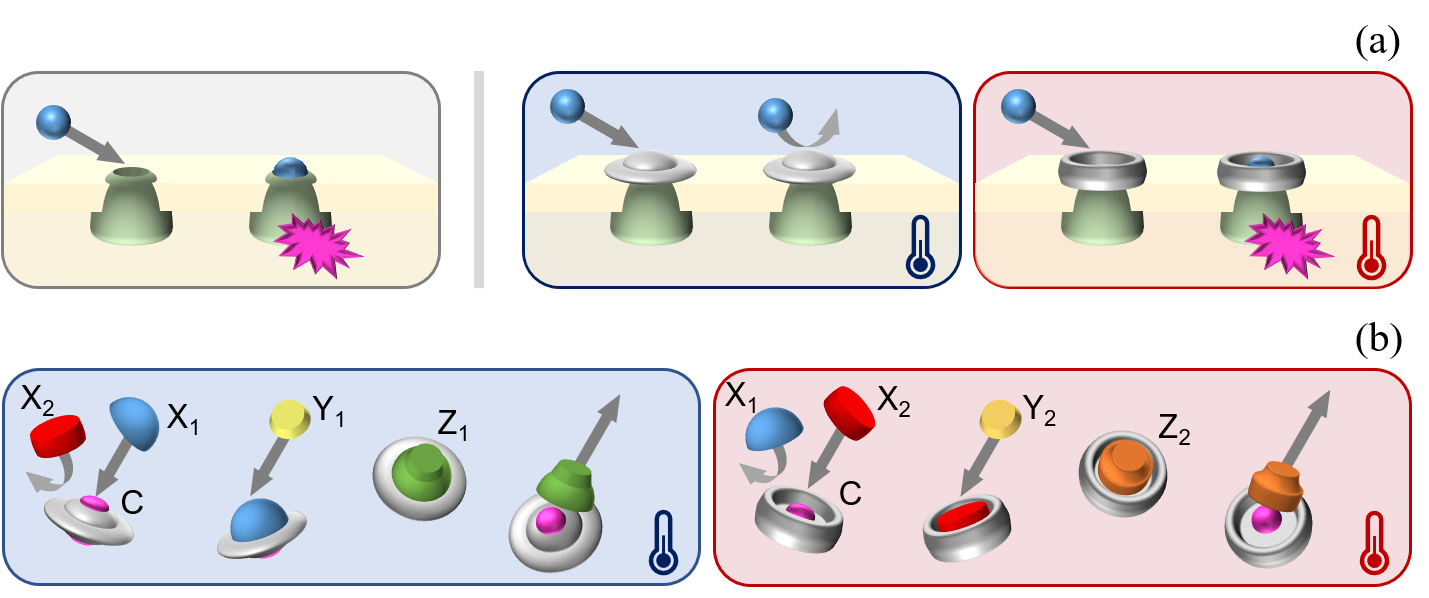}
\caption{
(a) Left: An agonist binds to 
the receptor.
Middle: At $T=0.1$, agonists are blocked by the EQM on the receptor.
Right: At $T \simeq 3.23$, an agonist passes through the EQM and binds to the receptor.
(b) The EQM with the catalyst in it.
It catalyzes different reactions depending on the temperature.
}
\label{fig:application}
\end{figure*}

First, 
consider a receptor (protein molecule) 
and an agonist that triggers a physiological response 
by binding to a certain site of the receptor
[Fig.~\ref{fig:application}(a), left].

Suppose that we cover the site with the EQM. 
At low temperature $T=0.1$, 
the inner circle of the EQM has higher particle density, 
as shown in Fig.~\ref{fig:equilibrium} (c). 
Hence the EQM blocks the agonist
[Fig.~\ref{fig:application}(a), middle].
and the receptor is not activated.
However, 
we can raise the temperature of the EQM to $T \simeq 3.23$
by applying an external field or by raising the temperature of
the environment.
Then, the density profile in the EQM is altered 
in such a way that the inner circle has a lower density,
as shown in Fig.~\ref{fig:equilibrium} (c). 
Consequently, the agonist can pass through the EQM with a non-vanishing probability,
binds to the site, and the receptor is activated
[Fig.~\ref{fig:application}(a), right].

In this way, 
we can control the binding of the agonist by changing the temperature of the EQM.

\subsection{Switchable nanozyme}
Second,
consider a catalyst C, such as a metallic atom,
that  accelerates two kinds of reactions
$X_1+Y_1 \to Z_1$ 
and 
$X_2+Y_2 \to Z_2$.

Suppose that 
all the reactants 
are dissolved in a solvent,
but we want to choose which one of $Z_1$ or $Z_2$
is produced.
Our EQM makes this possible as follows.
Put 
C on the center of the EQM, 
as shown in Fig.~\ref{fig:application}(b).
At 
$T=0.1$, 
the particle density in the EQM has a characteristic pattern  
as shown in Fig.~\ref{fig:equilibrium} (c).
Hence, only the molecule $X_1$ that geometrically fits this 
pattern is catalyzed.
When the temperature of the EQM is raised to $T \simeq 3.23$,
the density profile in the EQM changes into a much different pattern, 
as Fig.~\ref{fig:equilibrium} (c). 
Then, the catalyzed reaction of $X_1$ is blocked, 
whereas the reaction of $X_2$ that fits the new geometric pattern is catalyzed.

Thus, the EQM works as a nanozyme that
catalyzes different reactions 
depending on the temperature of the EQM, 
and the temperature can be controlled by an external field
as well as by the temperature of the environment.

\section{Summary and discussion}

We have studied 
nanomachines which we call the entropic quantum machines.
Since the relevant degrees of freedom of the EQM is large 
but smaller than those of proteins, 
both the entropic effect and 
the quantum effect over the whole system play the essential roles
in producing nontrivial functions.
To design the EQMs, 
we have proposed a systematic protocol,
which is based on the recent progress of the pure-state quantum statistical mechanics.
As a demonstration, we have proposed and analyzed an EQM which 
exhibits two characteristic patters of distributions of particles
by changing temperature or by applying a pulsed external field.
The two patters can be switched reversibly and repeatably.
We have also discussed potential applications of this EQM.
Although a simple model is assumed for this EQM, 
we expect more practical and interesting EQMs can also be designed 
and analyzed using our protocol.

\begin{acknowledgments}
We thank M. Onaka, K. Asai, S. Hiraoka, H. Katsura, Y. Arakawa and M. Ueda for discussions. 
R.H. was supported by the
Japan Society for the Promotion of Science through Program
for Leading Graduate Schools (ALPS).
This work was supported by the Japan Society for the Promotion of Science, KAKENHI No. 15H05700 and No. 19H01810.
\end{acknowledgments}


\bibliography{HS2020_20200512.bbl}

\end{document}